\documentclass[11pt]{article}
\usepackage{color}
\usepackage[dvipdfmx]{graphicx}        
\begin{document}

\begin{center}{\Large The Information Criterion GIC \\
of the Trend and Seasonal Adjustment Models}\\

\vspace{5mm}
{\large Genshiro Kitagawa}\\
Mathematics and Informatics Center, The University of Tokyo\\
and Meiji Institute for Advanced Study of Mathematical Sciences, Meiji University

\vspace{3mm}
{\today}

\end{center}

\noindent{\bf Abstract}

This paper presents an algorithm for computing the GIC and the TIC of the nonstationary 
state-space models.
The gradient and Hessian of the log-likelihood neccesary in computing the GIC
are obtained the differential filter that is obtained by extending the Kalman filter.
Three examples of the nonstationary time series models, i.e., the trend model,
statndard seasonal adjustment model and the seasonal adjustment model with
stationary AR component are presented to
exemplified the specification of structural matrices.
\\


\section{Introduction: The Maximum Likelihood Estimation of a State-Space Model}

We consider a linear Gaussian state-space model 
\begin{eqnarray}
 x_n &=& F_n(\theta) x_{n-1} + G_n(\theta) v_{n} \label{Eq-SSM-1}\\
 y_n &=& H_n(\theta) x_n + w_n,  \label{Eq-SSM-2}
\end{eqnarray}
where $y_n$ is a one-dimensional time series, $x_n$ is an $m$-dimensional state vector, $v_n$ is a $k$-dimesional Gaussian white noise, $v_n \sim N(0,Q_n(\theta ))$, and $w_n$ is one-dimensional white noise, $w_n \sim N(0,R_n(\theta ))$.
$F_n(\theta )$, $G_n(\theta )$ and $H_n(\theta )$ are $m\times m$ matrix, $m\times k$ matrix and $m$ vector. respectively.
$\theta$ is the $p$-dimensional parameter vector of the state-space model such as the variances of the noise
inputs and unknown coefficients in the matrices $F_n(\theta )$, $G_n(\theta )$, $H_n(\theta )$, $Q_n(\theta )$ and $R_n(\theta )$.
For simplicity of the notation, hereafter, the parameter $\theta$ and the suffix $n$ will be omitted.

Various models used in time series analysis, such as the stationary AR and ARMA models, and
various nonstationary models including trend model and the seasonal adjustment model,
can be treated uniformly within the state-space model framework. 
Further, many problems of time series analysis, such as prediction, signal extraction, decompositoon, parameter estimation and interpolation, can be formulated as the state estimation of a state-space model.

Given the time series $Y_N\equiv \{y_1,\ldots ,y_N\} $ and the state-space model (\ref{Eq-SSM-1})  
and (\ref{Eq-SSM-2}), the one-step-ahead predictor $x_{n|n-1}$ and the filter $x_{n|n}$ and their variance covariance 
matrices $V_{n|n-1}$ and $V_{n|n}$ are obtained by the following Kalman filter (Anderson and Moore (2012) and Kitagawa (2020)):

One-step-ahead prediction
\begin{eqnarray}
 x_{n|n-1} &=& F x_{n-1|n-1} \nonumber \\
 V_{n|n-1} &=& F V_{n-1|n-1} F^T + G Q_{n} G^T \label{Eq-3-2}
\end{eqnarray}
\indent
Filter 
\begin{eqnarray}
 K_n &=& V_{n|n-1}H^T (H V_{n|n-1} H^T + R)^{-1} \nonumber \\
 x_{n|n} &=& x_{n|n-1} + K_n (y_n -H x_{n|n-1}) \label{Eq-3-3} \\
 V_{n|n} &=& (I -K_n H)V_{n|n-1}. \nonumber
\end{eqnarray}

Given the data $Y_N$, the likelihood of the time series model is defined by 
\begin{eqnarray}
 L(\theta ) &=& p( Y_N|\theta ) 
  =  \prod_{n=1}^N g_n(y_n|Y_{n-1},\theta ),
\end{eqnarray}
where $g_n(y_n|Y_{n-1},\theta )$ is the conditional distribution of $y_n$ given the observation $Y_{n-1}$ and is a normal distribution given by
\begin{eqnarray}
g_n(y_n|Y_{n-1},\theta ) 
  = \frac{1}{\sqrt{2\pi r_n}}\exp\left\{ -\frac{\varepsilon_{n}^2}{2r_n} \right\}, 
\label{eq_distrribution_g}
\end{eqnarray}
where $\varepsilon_n$ and $r_n$ are the one-step-ahead prediction error and
its variance defined by
\begin{eqnarray}
\varepsilon_n &=& y_n - Hx_{n|n-1} \nonumber \\
r_n &=&  H_n V_{n|n-1} H_n^T + R.  \label{Eq_prediction_error}
\end{eqnarray}

Therefore, the log-likelihood of the state-space model is obtained as
\begin{eqnarray}
 \ell (\theta ) = \log L(\theta ) &=& \sum_{n=1}^N \log g_n(y_n|Y_{n-1},\theta ) \nonumber \\
  &=& -\frac{1}{2} \biggl\{ N \log 2\pi 
      + \sum_{n=1}^N \frac{\varepsilon_n^2}{r_n} + \sum_{n=1}^N \log r_{n}
  \biggr\}. \label{Eq_log-lk}
\end{eqnarray}

The maximum likelihood estimates of the parameters of the state-space model 
can be obtained by maximizing the log-likelihood function.
In general, since the log-likelihood function is mostly nonlinear, the maximum likelihood estimates  is obtained by using a numerical optimization algorithm based on the quasi-Newton method. 
According to this method, using the value  $\ell( \theta)$ of the log-likelihood and the first derivative (gradient) $\partial \ell/\partial \theta$ for a given parameter \( \theta \), the maximizer of \( \ell( \theta) \) is automatically estimated by repeating 
\begin{equation}
 \theta_k = \theta_{k-1} + \lambda_k B_{k-1}^{-1} \frac{\partial \ell}{\partial \theta},
\end{equation}
where $\theta_0$ is an initial estimate of the parameter. 
The step width \( \lambda_k \) is automatically determined and the inverse matrix \( H_{k-1}^{-1} \) of the Hessian matrix is obtained recursively by the DFP or BFGS algorithms (Fletcher (2013)).

Here, the gradient of the log-likelihood function is usually approximated by numerical difference,
such as 
\begin{eqnarray}
\frac{\partial \ell (\theta)}{\partial \theta_j} \approx \frac{\ell (\theta_j +\Delta \theta_j) -  \ell (\theta_j  \Delta \theta_j)}{2\Delta\theta_j},
\end{eqnarray}
where $\Delta\theta_j$ is defined by $C |\theta_j|$, for some small $C$ such as 0.0001.
The numerical difference usually yields reasonable approximation to the gradient of the 
log-likelihood.
However, since it requires $2p$ times of log-likelihood evaluations, the amount of computation
becomes considerable if the dimension of the parameters is large.
Further, if the the maximum likelihood estimates lie very close to the boundary of
addmissible domain, which sometimes occure in regularization problems,
it becomes difficlt to obtain the approximation to the gradient of the log-likelihood
by  the numerical difference.

Analytic derivative of the log-likelihood of time series models were considered by many authors. 
For example,  Kohn and Ansley (1985) gave method for computing likelihood and its derivatives for an ARMA model.
Zadrozny (1989) derived analytic derivatives for estimation of linear dynamic models.
Kulikova (2009) presented square-root algorithm for the likelihood gradient evaluation to avoid numerical instatbility of the recursive algorithm for log-likelihood computation.
In Kitagawa (2021,2022), algorithms for computing the gradient and Hessian of the log-likelihood 
of linear state-space model are given. 
Details of the implementation of the algorithm for standard seasonal adjustment model, seasonal adjustment model with stationary AR component and ARMA model are given.
For each implementation, comparison with a numerical difference method is shown.

In the state-space modeling of time series, the evaluation of the estimated models isimportant.
For that purpose, the information criterion AIC is a standard method for the model evaluation and
model comparison.
The information criterion GIC was derived as a genenral model evaluation criterion which can be 
applied to not only the models whose parameters are estimated by the maximum likelihood method,
but also a broad clasee of estimators defined by statistical functionals, that cpvers various 
types of regularization methods and some type of Bayesian models (Konishi and Kitagawa (1996, 2008).
Although GIC is theoretically appealing, the difficulty with the GIC in applying the time series
model is that it is difficult to compute the GIC since it is necessary to compute the Hessian of
the log-likelihood.
In this paper, we show the application of the differential filter for computing the GIC of the 
time series models that can be expressed by a state-space model.

In section 2, we briefly present the differential filter to obtain the gradient and the Hessian 
of the log-likelihood of the state-space model (Kitagawa (2021)).
The formulas for computing the Fisher information matrix and the Hessian of the log-likelihood
that are necessary for computing the GIC are also explained in this section.
In section 3, three noonstationary time series models. the trend model, the standard seasonal adjustment model
and the seasonal adjustment model with stationary AR component are shown to 
exemplify the model.
In the apllication of the differential filter, it is necessary to specify the first and second
order derivative of the structural parameters of the state-space models, i.e., the matrices 
$F$, $G$, $H$, $Q$ and $R$.
In the examples, it will be shown that the most of these terms zero or at least very sparse
that makes the computation of the gradient filter rather simple.


\section{The Gradient and the Hessian of the log-likelihood }
\subsection{The gradient of the log-likelihood}

The general formuara of the differential filter for computing the gradient and the Hessian
of the log-likelihood is very complex (Kitagawa 2021,2022).
However, for the time series models considered in section 3, the derivatives of the matrix
$F$, $G$, $H$, $Q$ and $R$ satisfy
\begin{eqnarray}
 \frac{\partial^2 F}{\partial\theta\partial\theta^T}=0,\quad 
 \frac{\partial G}{\partial\theta}=0,\quad 
 \frac{\partial^2 G}{\partial\theta\partial\theta^T}=0,\quad 
 \frac{\partial H}{\partial\theta}=0,\quad 
 \frac{\partial^2 H}{\partial\theta\partial\theta^T}=0,\nonumber 
\end{eqnarray}
where the zero in the right hand side is the zero matrix 
that makes the algorithm of the differential filter fairly simple.
Further, if $\displaystyle \frac{\partial F}{\partial\theta}=0$ as is the cases 
for examples shown in subsection 3.1 and 3.2, in the following algorithms, the terms written in
red disappear, which make the computation much simpler.

From (\ref{Eq_log-lk}), the gradient of the log-likelihood is obtained by
\begin{eqnarray}
\frac{\partial\ell (\theta )}{\partial \theta} 
&=& - \frac{1}{2} \sum_{n=1}^N  \left( 
   \frac{1}{r_n}\frac{\partial r_n}{\partial\theta}
  +  2\frac{\varepsilon_n}{r_n}\frac{\partial \varepsilon_n}{\partial\theta}
  -  \frac{\varepsilon_n^2}{r_n^2}\frac{\partial r_n}{\partial\theta}
   \right),  \label{Eq_gradient_ell}
\end{eqnarray}
where, from (\ref{Eq_prediction_error}), the derivatives of the one-step-ahead predition $\varepsilon_n$ and the
one-step-ahead prediction error variance $r_n$ are obtained by
\begin{eqnarray}
\frac{\partial \varepsilon_n}{\partial\theta} &=& -H \frac{\partial x_{n|n-1}}{\partial\theta}
  \label{Eq_gradient_pred_error}\\
\frac{\partial r_n}{\partial\theta} &=& H \frac{\partial V_{n|n-1}}{\partial\theta}H^T
   + \frac{\partial R}{\partial\theta}. \label{Eq_gradient_observation_error}  
%
\label{Eq_gradient_observation_noise_variance}
\end{eqnarray}

To evaluate these quantity, we need the derivative of the one-step-ahead predictor
of the state $\displaystyle\frac{\partial x_{n|n-1}}{\partial\theta}$ and its variance covariance
matrix $\displaystyle\frac{\partial V_{n|n-1}}{\partial\theta}$ which can be obtained recursively
in parallel to the Kalman filter algorithm:\\

\quad [One-step-ahead-prediction] \\
\begin{eqnarray}
\frac{\partial x_{n|n-1}}{\partial\theta} &=& F \frac{\partial x_{n-1|n-1}}{\partial\theta}
  {\color{red} + \frac{\partial F}{\partial\theta} x_{n-1|n-1} }
\nonumber \\
\frac{\partial V_{n|n-1}}{\partial\theta} &=& F \frac{\partial V_{n-1|n-1}}{\partial\theta}F^T 
      {\color{red}     + \frac{\partial F}{\partial\theta}V_{n-1|n-1}F^T 
           + F V_{n-1|n-1} \frac{\partial F}{\partial\theta}^T  }  
           + G\frac{\partial Q}{\partial\theta}G^T 
.  \label{Eq_gradient-filter-P}
\end{eqnarray}

\quad [Filter] \\
\begin{eqnarray}
\frac{\partial K_n}{\partial\theta} 
        &=&   
          \frac{\partial V_{n|n-1}}{\partial\theta}H^T  r_n^{-1} 
          - V_{n|n-1}H^T \frac{\partial r_n}{\partial\theta}r_n^{-2 }    \nonumber \\
\frac{\partial x_{n|n}}{\partial\theta} &=&  
        \frac{\partial x_{n|n-1}}{\partial\theta}
       + K_n \frac{\partial \varepsilon_n}{\partial\theta}
       + \frac{\partial K_n}{\partial\theta} \varepsilon_n   \nonumber \\
\frac{\partial V_{n|n}}{\partial\theta} &=& 
         \frac{\partial V_{n|n-1}}{\partial\theta}
       - \frac{\partial K_n}{\partial\theta} H V_{n|n-1}
       - K_n H \frac{\partial V_{n|n-1}}{\partial\theta}. \label{Eq_gradient-filter-F}
\end{eqnarray}


\subsection{Hessian of the Log-likelihood of the State-space Model}

To compute the GIC of the model, it is necessary to obtain 
the Hessian (the second derivative) of the log-likelihood which can 
also be obtained by a recursive formula, 
since, from (\ref{Eq_gradient_ell}), it is given as
\begin{eqnarray}
\frac{\partial^2\ell (\theta )}{\partial \theta\partial \theta^T}
&=& - \frac{1}{2}\sum_{n=1}^N \left\{ 
   \frac{1}{r_n}\left( 
    \frac{\partial^2 r_n}{\partial\theta\partial\theta^T} 
  +2\frac{\partial \varepsilon_n}{\partial\theta}\frac{\partial \varepsilon_n}{\partial\theta^T}  
  +2\varepsilon_n \frac{\partial^2 \varepsilon_n}{\partial\theta \partial\theta^T}
  \right) 
  -\frac{1}{r_n^2}\left(
   \frac{\partial r_n}{\partial\theta}\frac{\partial r_n}{\partial\theta^T}
 \right. \right.   \nonumber \\
&& {}\qquad  \left. \left. 
+ 2\varepsilon_n \frac{\partial r_n}{\partial\theta}\frac{\partial \varepsilon_n}{\partial\theta^T}
+ 2\varepsilon_n \frac{\partial \varepsilon_n}{\partial\theta}\frac{\partial r_n}{\partial\theta^T}
+ \varepsilon_n^2\frac{\partial^2 r_n}{\partial\theta \partial\theta^T} \right)
+ \frac{\varepsilon_n^2}{r_n^3}\frac{\partial r_n}{\partial\theta }\frac{\partial r_n}{\partial\theta^T}  \right\}
, \nonumber
\end{eqnarray}
where, from (\ref{Eq_gradient_observation_error}), 
$ \displaystyle\frac{\partial^2 \varepsilon_n}{\partial\theta\partial\theta^T} $ 
and $ \displaystyle\frac{\partial^2 r_n}{\partial\theta \partial\theta^T} $
 are
obtained by
\begin{eqnarray}
\frac{\partial^2 \varepsilon_n}{\partial\theta\partial\theta^T} 
&=&  -  H\frac{\partial^2 x_{n|n-1}}{\partial\theta\partial\theta^T}
  \\
\frac{\partial^2 r_n}{\partial\theta \partial\theta^T} 
   &=& H \frac{\partial^2 V_{n|n-1}}{\partial\theta \partial\theta^T}H^T  
      + \frac{\partial^2 R}{\partial\theta\partial\theta^T}. \nonumber 
\end{eqnarray}

Therefore, to evaluate the Hessian, the following computation should be performed
along with the recursive formula for the log-likelihood and the 
gradient of the log-likelihood.%

\begin{eqnarray}
\frac{\partial^2 x_{n|n-1}}{\partial\theta \partial\theta^T}
   &=&  F \frac{\partial^2 x_{n-1|n-1}}{\partial\theta \partial\theta^T}
       {\color{red} +  \frac{\partial F}{\partial\theta^T} \frac{\partial x_{n-1|n-1}}{\partial\theta}
        + \frac{\partial F}{\partial\theta} \frac{\partial x_{n-1|n-1}}{\partial\theta}  }
\nonumber \\
\frac{\partial^2 V_{n|n-1}}{\partial\theta \partial\theta^T}
         &=&   F \frac{\partial^2 V_{n-1|n-1}}{\partial\theta \partial\theta^T}F^T  
     {\color{red}  + \frac{\partial F}{\partial\theta^T} \frac{\partial V_{n-1|n-1}}{\partial\theta}F^T  
           + \frac{\partial F}{\partial\theta} \frac{\partial V_{n-1|n-1}}{\partial\theta^T}F^T  
         + F \frac{\partial V_{n-1|n-1}}{\partial\theta^T}\frac{\partial F^T}{\partial\theta}  }  \nonumber \\
       && {\color{red}+ F \frac{\partial V_{n-1|n-1}}{\partial\theta}\frac{\partial F^T}{\partial\theta^T} 
          + \frac{\partial^2 F}{\partial\theta \partial\theta^T}V_{n|n-1}F^T 
          + \frac{\partial F}{\partial\theta^T}V_{n|n-1}\frac{\partial F^T}{\partial\theta} 
          + \frac{\partial F}{\partial\theta}V_{n|n-1}  \frac{\partial F^T}{\partial\theta^T}  } \nonumber \\
       && {\color{red} + F V_{n|n-1} \frac{\partial^2 F^T}{\partial\theta \partial\theta^T} 
          + \frac{\partial G}{\partial\theta^T} \frac{\partial Q}{\partial\theta}G^T 
          + \frac{\partial G}{\partial\theta} \frac{\partial Q}{\partial\theta^T}G^T 
          + G\frac{\partial Q}{\partial\theta^T}\frac{\partial G^T}{\partial\theta}   }
          + G\frac{\partial^2 Q}{\partial\theta \partial\theta^T}G^T \nonumber \\
       && {\color{red} +  G\frac{\partial Q}{\partial\theta}\frac{\partial G^T}{\partial\theta^T}  
          + \frac{\partial^2 G}{\partial\theta\partial\theta^T}Q G^T 
          + \frac{\partial G}{\partial\theta^T}Q \frac{\partial G^T}{\partial\theta} 
          + \frac{\partial G}{\partial\theta}Q \frac{\partial G^T}{\partial\theta^T} 
          + G Q \frac{\partial^2 G^T}{\partial\theta \partial\theta^T}  }  \nonumber \\
\frac{\partial^2 K_n}{\partial\theta \partial\theta^T} 
        &=& \frac{\partial^2 V_{n|n-1}}{\partial\theta \partial\theta^T}H^T  r_n^{-1} 
           - \left( \frac{\partial V_{n|n-1}}{\partial\theta^T}H^T  
              \frac{\partial r_n}{\partial\theta}
              +  V_{n|n-1}H^T  \frac{\partial^2 r_n}{\partial\theta \partial\theta^T}  
           \right) r_n^{-2}  \nonumber \\
        &&{}\qquad  + 2V_{n|n-1}H^T 
            \frac{\partial r_n}{\partial\theta} \frac{\partial r_n}{\partial\theta^T} r_n^{-3} \\
\frac{\partial^2 x_{n|n}}{\partial\theta \partial\theta^T} &=&  
        \frac{\partial^2 x_{n|n-1}}{\partial\theta \partial\theta^T}
       +  \frac{\partial K_n}{\partial\theta} \frac{\partial \varepsilon_n}{\partial\theta^T}
       + \frac{\partial K_n}{\partial\theta^T} \frac{\partial \varepsilon_n}{\partial\theta}  
       + K_n \frac{\partial^2 \varepsilon_n}{\partial\theta \partial\theta^T} 
       + \frac{\partial^2 K_n}{\partial\theta \partial\theta^T} \varepsilon_n
\nonumber \\
\frac{\partial^2 V_{n|n}}{\partial\theta \partial\theta^T} &=& 
         \frac{\partial^2 V_{n|n-1}}{\partial\theta \partial\theta^T}
       - \frac{\partial^2 K_n}{\partial\theta \partial\theta^T} H V_{n|n-1}
       - \frac{\partial K_n}{\partial\theta^T} H \frac{\partial V_{n|n-1}}{\partial\theta} 
       - \frac{\partial K_n}{\partial\theta}H \frac{\partial V_{n|n-1}}{\partial\theta^T}
\nonumber \\
    && - K_n H \frac{\partial^2 V_{n|n-1}}{\partial\theta \partial\theta^T}. \nonumber
\end{eqnarray}

\subsection{Information Criteria GIC for the State-Space Model}

The information criterion GIC\cite{KK 2008} for the state-space model is given by
\begin{eqnarray}
   {\rm GIC} = -2 \log L(\hat\theta ) 
              + 2 \mbox{tr} \left( I(\hat\theta )J(\hat\theta )^{-1} \right),
\end{eqnarray}
where $\hat\theta$ is the maximum likelihood estimate of the parameter $\theta$, and $I(\hat\theta )$ and $J(\hat\theta )$ are the Fisher information and negative of the Hessian defined by
\begin{eqnarray}
I(\hat\theta ) &=& \frac{1}{N} \sum_{j=1}^N \left. \frac{\partial \ell (\theta)}{\partial\theta} \frac{\partial \ell (\theta)}{\partial\theta^T}\right|_{\theta =\hat\theta} \nonumber \\
J(\hat\theta ) &=& - \frac{1}{N} \sum_{j=1}^N \left. \frac{\partial^2 \ell(\theta)}{\partial\theta\partial\theta^T} \right|_{\theta =\hat\theta}.
\end{eqnarray}
These term can be obtained as the byproduct of the differential filter.
Note that if the parameters are estimated by the maximum likelihood method,
GIC is identical to the TIC\cite{KK 2008}\cite{Takeuchi 1976}.

\section{Examples}
In order to impliment the differential filter, it is necessary to to specify 
the first and the second derivatives
of $F$, $G$, $H$, $Q$ and $R$ along with the original state-space model.
In this section, we shall consider three typical cases.
The first two examples are the trend model and the standard seasonal adjeustment model, for which three matrics (or
vector), $F$, $G$ and $H$ do not contain unknown parameters and thus the derivatives
of these matrics becomes 0. 
This makes the algorithm for the gradient and the Hessian of the log-likelihood considerablly simple.
The third example is the seasonal adjustment model with AR component.
For this model,  the matrix $F$ depends on the unknown AR coefficients, 
but the derivative of $F$ is very simple and very sparse.

\subsection{Trend model}

This is a typical example of the case where only the noise covariances $Q$ and $R$
depend on the unknown parameter $\theta$.
Consider a trend model
\begin{eqnarray}
y_n = T_n + w_n,
\end{eqnarray}
where $T_n$ is the trend component 
that typically follow the following model
\begin{eqnarray}
(1-B)^k T_n = v_n, \label{eq_trend_model}
\end{eqnarray}
where $B$ is the back-shift operator satisfying $BT_n = T_{n-1}$,
 $v_n$ and $w_n$ are assumed to be Gaussian white noise with
variances $\tau^2$ and $\sigma^2$, respectively (Kitagawa and Gersch (1984,1996) and 
Kitagawa (2020)).
Note that for $k=1$ and $k=2$, the model (\ref{eq_trend_model}) becomes
$T_n = T_{n-1}+v_n$ and $T_n = 2T_{n-1} - T_{n-2} + v_n$, respectively.

This trend model can be expressed in
the state-space model form as
\begin{eqnarray}
x_n &=& F x_{n-1} + G v_n \nonumber \\
y_n &=& H x_n + w_n,
\end{eqnarray}
with $v_n \sim N(0,Q)$ and $w_n \sim N(0,R)$ and the state vector $x_n$ and 
the matrices $F$, $G$, $H$, $Q$ and $R$ are defined by
\begin{eqnarray}
x_n &=& T_n,\quad  F = 1 ,\quad G = 1, \quad H =1, \quad
Q = \tau^2, \quad  R = \sigma^2,
\end{eqnarray}  
for $k=1$ and
\begin{eqnarray}
x_n &=& \left[ \begin{array}{c} 
             T_n     \\
             T_{n-1}  \end{array}\right],\quad
F = \left[ \begin{array}{cc} 
             2 &-1  \\
             1 & 0  \end{array}\right],\quad
G = \left[ \begin{array}{cc} 
             1 & 0   \\
             0 & 0   \end{array}\right] \\
H &=& [\begin{array}{cc} 1 &0 \end{array}], \quad
Q = \tau^2, \quad  R = \sigma^2,
\end{eqnarray}  
for $k=2$, respectively.

In this state-space representation, the parameter is $\tau^2$,
and the $F$, $G$ and $H$ do not depend on the parameter.
Therefore, we have $\displaystyle\frac{\partial F}{\partial\theta}=
\displaystyle\frac{\partial G}{\partial\theta}=\displaystyle\frac{\partial H}{\partial\theta}=0$
and $\displaystyle\frac{\partial^2 F}{\partial\theta\partial\theta^T}=
\displaystyle\frac{\partial^2 G}{\partial\theta\partial\theta^T}=
\displaystyle\frac{\partial^2 H}{\partial\theta\partial\theta^T}=0$.

In actual likelihood maximization, since there are positivity 
constraints, $\tau^2 > 0$, and $\sigma^2 >0$,
it is frequently used log-transformations,
\begin{eqnarray}
 \theta_1 = \log (\tau^2) ,\quad \theta_2 = \log( \sigma^2 )
\end{eqnarray}
and maximize the log-likelihood with respect to this transformed parameter
 $\theta = (\theta_1,\theta_2)^T$.
In this case, 
\begin{eqnarray}
  \frac{\partial Q}{\partial\theta_1} =
  \frac{\partial^2 Q}{\partial\theta_1 \partial\theta_1} = 
     \left[ \begin{array}{cc} \tau^2 & 0 \\
                          0   & 0 \end{array}\right], \quad
  \frac{\partial Q}{\partial\theta_2} =
  \frac{\partial^2 Q}{\partial\theta_1 \partial\theta_2} = 
  \frac{\partial^2 Q}{\partial\theta_2 \partial\theta_1} = 
  \frac{\partial^2 Q}{\partial\theta_2 \partial\theta_2} = 
     \left[ \begin{array}{cc} 0 & 0 \\
                              0 & 0 \end{array}\right], \nonumber \\
  \frac{\partial R}{\partial\theta_2} =
  \frac{\partial^2 R}{\partial\theta_2 \partial\theta_2} =
        \left[ \begin{array}{cc} \sigma^2 & 0 \\
                          0   & 0 \end{array}\right], \quad
  \frac{\partial R}{\partial\theta_2} =
  \frac{\partial^2 R}{\partial\theta_1 \partial\theta_1} = 
  \frac{\partial^2 R}{\partial\theta_1 \partial\theta_2} = 
  \frac{\partial^2 R}{\partial\theta_2 \partial\theta_1} =
        \left[ \begin{array}{cc} 0 & 0 \\
                              0 & 0 \end{array}\right]  .
\end{eqnarray}
Since log-transfomation is a monotone incresing function, we can get the 
same parameter by solving this modified optimization problem.

\begin{table}[bp]
\begin{center}
\caption{Comparison of numerical diffference and gradient methods for the first order trend model.}\label{Tab_Trend_model_m1=1}

\vspace{2mm}
\begin{tabular}{c|cc}
        & Initial Model &  MLE \\
\hline
  $(\tau^2, \sigma^2)$ &  $ [ \begin{array}{cc} 0.1\!\times \!\!10^{-3}  & 0.2\!\times \!\!10^{-3} \end{array} ]$ &      $ [ \begin{array}{cc} 6.87264\!\times \!\!10^{-4}  & 1.31613\!\times \!\!10^{-4} \end{array} ]$\\[1mm]
  $\theta$   & $ [ \begin{array}{cc} -9.21034  & -8.51719 \end{array} ]$ &
      $ [\begin{array}{cc} -7.28279  &  -8.93564 \end{array} ]$\\[1mm]
  $\ell (\theta )$ & 253.1868  &  317.5342 \\[1mm]
  $\displaystyle\frac{\partial \ell(\theta )}{\partial \theta}$ & $\left[\begin{array}{cc} 72.417363& 59.11161 \end{array} \right]$ & $ [\begin{array}{cc}0.1347\!\times \!\!10^{-8}& -1.2742\!\times \!\!10^{-8}\end{array}]$ \\[3mm]
  $\displaystyle\frac{\partial^2 \ell(\theta )}{\partial \theta\partial\theta^T}$ & 
  $\left[ \begin{array}{cc} 35.77478 & 62.19858\\
                            62.19858 & 48.28391 \end{array}\right] $  &
  $\left[ \begin{array}{rr} 45.69891 & 12.22819\\
                            12.22819 &  6.84511 \end{array}\right] $ \\[3mm]
  $b(\mbox{GIC})$ & -2.4551 &  1.4547   \\
\hline
\end{tabular}
\end{center}
\end{table}

\vspace{3mm}
For Whard (whole sale hardware) data (Kitagawa (2020)), $N=155$, 
the parameter $\theta = (\log\tau^2,\log\sigma^2)^T$ of the trend model with $m_1=1$ was estimated 
using the initial values $\tau^2_0 = 0.1\times 10^{-3}$ and $\sigma^2_0 = 0.2\times 10^{-3}$. 
By a numerical optimization procedure, the maximum likelihood estimates of the parametesr are obtained
 as  $\hat\tau^2 = 0.687264\times 10^{-3} $ and $\hat\sigma^2 =0.131613\times 10^{-3}$.
In this case, the bias correction term of the GIC, $I(\hat\theta)J(\hat\theta)^{-1}$ is
evaluated as 1.4547. Note that since this model contains two parameters, the bias correction
terms is 2.

Table \ref{Tab_Trend_model_m1=1} shows the log-likelihoods, the gradients, the Hessians and the 
observation noise variances of the initial and the final estimates.
The log-likelihood of the model with these initial and final estimates are 
$\ell (\theta )= -307.6616$ and $-317.9243$, respectively.



Table \ref{Tab_Trend_model_m1=2} shows the results for the sencond order trend model.
In this case, the final estimate obtained by the numerical optimization procedure
depends on the initial estimate and two cases are shown in the table.
If the inital estiamte is set to $\tau^2_0=10^{-4}$ and $\sigma^2_0=2\times 10^{-4}$, 
the final estimate is $\hat\tau^2=1.9222\times 10^{-4}$ and $\hat\sigma^2 = 3.4960\times 10^{-4}$
with the log-likelihood value $\ell (\hat\theta )=293.0193$.
On the other hand, if we set the initial estimate as $\theta_0=2\times 10^{-7}$ and 
$\sigma_0^2 = 2\times 10^{-4}$, the final estiamte becomes $\hat\theta = 1.16423\times 10^{-6}$
and $\hat\sigma^2 = 1.11047\times 10^{-1}$ with $\ell (\hat\theta )=278.6631$.
Comparing the log-likelihodd values, $\hat\tau^2=1.9222\times 10^{-4}$ and $\hat\sigma^2 = 3.4960\times 10^{-4}$
are the maximum likelihood estimate of the second order trend model.
The bias correction term $b$(GIC) is evaluated as 1.9115 and 2.8151, respectively.
Note that the bias correction term $b$(GIC) takes different values depending on the 
estimated parameters.
This is because the curveture of the log-likelihood function is different dependeing on the 
pair of parameter values.

\begin{table}[tbp]
\begin{center}
\caption{Comparison of numerical diffference and gradient methods for the second order trend model.}\label{Tab_Trend_model_m1=2}

\vspace{2mm}
\begin{tabular}{c|cc}
        & Initial model &  Optimized model \\
\hline
  $(\tau^2, \sigma^2)$ & $ [ \begin{array}{cc} 0.1\times 10^{-3}  & 0.2 \times 10^{-3} \end{array} ]$ &
     $ [\begin{array}{cc} 1.9222\!\times \!\!10^{-4}  &  3.4960\!\times\!\! 10^{-4} \end{array} ]$\\[1mm]
  $\theta$ & $ [ \begin{array}{cc} -9.21034  & -8.51719 \end{array} ]$ & 
     $ [\begin{array}{cc} -8.55687  &  -7.95871 \end{array} ]$  \\[1mm]
  $\ell (\theta )$ & 276.6621  &  293.0193 \\[1mm]
  $\displaystyle\frac{\partial \ell(\theta )}{\partial \theta}$ & $\left[\begin{array}{cc} 20.50334& 40.91088 \end{array} \right]$ & $ [\begin{array}{cc}0.8543\!\times \!\!10^{-8}& 0.3189\!\times \!\!10^{-8}\end{array}]$ \\[3mm]
  $\displaystyle\frac{\partial^2 \ell(\theta )}{\partial \theta\partial\theta^T}$ & 
  $\left[ \begin{array}{cc} 20.09159 & 24.63536\\
                            24.63536 & 68.55278 \end{array}\right] $  &
  $\left[ \begin{array}{rr} 14.27777 &  10.60807\\
                            10.60807 & 41.00773 \end{array}\right] $ \\[3mm]
  $b(\mbox{GIC})$ & 5.4927 &  1.9115  \\
\hline
  $(\tau^2, \sigma^2)$ & $ [ \begin{array}{cc} 0.2\times 10^{-6}  & 0.2 \times 10^{-3} \end{array} ]$ &
     $ [\begin{array}{cc} 1.16423\!\times \!\!10^{-6}  &  1.11047\!\times\!\! 10^{-1} \end{array} ]$\\[1mm]
  $\theta$ & $ [ \begin{array}{cc} -15.42495  & -8.51719 \end{array} ]$ & 
     $ [\begin{array}{cc} -8.55687  &  -7.95871 \end{array} ]$  \\[1mm]
  $\ell (\theta )$ & 37.3252  &  278.6631 \\[1mm]
  $\displaystyle\frac{\partial \ell(\theta )}{\partial \theta}$ & $\left[\begin{array}{cc} 20.46039& 328.96519 \end{array} \right]$ & $ [\begin{array}{cc}0.01318\!\times \!\!10^{-7}& \!-1.24767\!\times \!\!10^{-7}\end{array}]$ \\[3mm]
  $\displaystyle\frac{\partial^2 \ell(\theta )}{\partial \theta\partial\theta^T}$ & 
  $\left[ \begin{array}{cc} 4.54019 & 20.31043\\
                            20.31043 & 380.76459 \end{array}\right] $  &
  $\left[ \begin{array}{rr} 1.78678 & 2.66212\\
                            2.66212 & 69.38934 \end{array}\right] $ \\[3mm]
  $b(\mbox{GIC})$ & 37.3252 &  2.8151  \\
\hline
\end{tabular}
\end{center}
\end{table}

\subsection{The standard seasonal adjustment model}

As the second example, we consider a standard seasonal adjustment model
\begin{eqnarray}
y_n = T_n + S_n + w_n,
\end{eqnarray}
where $T_n$ and $S_n$ are the trend component and the seasonal component
that typically follow the following model
\begin{eqnarray}
&&T_n = 2T_{n-1} - T_{n-2} + u_n, \nonumber \\
&&S_n =-(S_{n-1}+\cdots +S_{n-p+1}) + v_n.
\end{eqnarray}
Theree noise terms, $u_n$, $v_n$ and $w_n$ are assumed to be Gaussian white noise with
variances $\tau_1^2$, $\tau_2^2$ and $\sigma^2$, respectively (Kitagawa and Gersch (1984,1996) and 
Kitagawa (2020)).

This seasonal adjustment model with two component models can be expressed in
state-space model form as
\begin{eqnarray}
x_n &=& F x_{n-1} + G v_n \nonumber \\
y_n &=& H x_n + w_n
\end{eqnarray}
with $v_n \sim N(0,Q)$ and $w_n \sim N(0,R)$ and the state vector $x_n$ and 
the matrices $F$, $G$, $H$, $Q$ and $R$ are defined by
{\setlength{\arraycolsep}{1mm}
\begin{eqnarray}
x_n = \left[ \begin{array}{c} 
             T_n     \\
             T_{n-1} \\
             S_n     \\
             S_{n-1} \\
            \vdots\\ 
             S_{n-p+2} \end{array}\right],\quad
F &=& \left[ \begin{array}{cccccc} 
             2 &-1 &   &   &   &   \\
             1 & 1 &   &   &   &   \\
               &   &-1 &-1 &\cdots&-1\\
               &   & 1 &   &   &   \\
               &   &   &\ddots&&   \\ 
               &   &   &   & 1 &    \end{array}\right],\quad
G = \left[ \begin{array}{cc} 
             1 & 0   \\
             0 & 0  \\
             0 & 1  \\
             0 & 0  \\
          \vdots&\vdots\\ 
             0 & 0   \end{array}\right] \\
H &=& [\begin{array}{cccccc} 1&0&1&0&\cdots &0 \end{array}] \nonumber \\
Q &=& \left[ \begin{array}{cc} 
             \tau_1^2 & 0   \\
               0      & \tau_2^2 \end{array}\right], \quad
R = \sigma^2.
\end{eqnarray}  
}

\begin{table}[bp]
\begin{center}
\caption{Comparison of numerical diffference and gradient}\label{Tab_TS_model}

\vspace{2mm}
\begin{tabular}{c|cc}
        & Initial model &  Optimal model \\
\hline
  $(\tau^2_1,\tau^2_2,\sigma^2)$ & $ [ \begin{array}{ccc}0.1\!\times \!\!10^{-3} & 0.2\!\times \!\!10^{-4} & \!0.2\!\times \!\!10^{-3} \end{array} ]$ &
  $ [ \begin{array}{ccc}0.55592\!\times \!\!10^{-5} & \!0.43372\!\times \!\!10^{-4} & 0.52734\!\times \!\!10^{-4} \end{array} ]$\\
  $\theta$   & $ [ \begin{array}{ccc}-9.21034 & \!-10.81978 & \!-8.51719 \end{array} ]$  & 
  $[ \begin{array}{ccc} -12.10001 & -10.04570 & -9.85025 \end{array} ]$   \\
  $\ell (\theta )$ & 346.5115  &  384.9600 \\[2mm]
  $\frac{\partial \ell(\theta )}{\partial \theta}$ & $\left[ -18.12229, -4.82792, -17.81465 \right]$ & $[-0.47689\!\times \!\!10^{-6}, -1.31732\!\times \!\!10^{-6}, 0.0856\!\times \!\!10^{-6}]$ \\[2mm]
  $\frac{\partial^2 \ell(\theta )}{\partial \theta\partial\theta^T}$ & 
  $\left[ \begin{array}{ccc} 5.77960 & -0.06331 & -1.89624\\
                            -0.06331 &  3.88402 &  2.54243\\
                            -1.89624 & 2.54243  & 20.14142\end{array}\right] $  &
  $\left[ \begin{array}{ccc} 8.66117 & 1.12346  & 3.41325\\
                             1.12346 & 18.99017 & 11.33307\\
                             3.41325 & 11.3307 & 11.97043\end{array}\right] $ \\[2mm]
  $b(GIC)$ & 1.1946 &  3.9558   \\
\hline
\end{tabular}
\end{center}
\end{table}

In this case, the parameter is $\theta =(\tau_1^2,\tau_2^2,\sigma^2)^T$,
and the $F$, $G$, $H$ and $R$ do not depend on the parameter.
In actual likelihood maximization, since there are positivity 
constrains, $\tau_1^2 > 0$, $\tau_2^2 >0$ and $\sigma^2 >0$,
we use the log-transformation,
\begin{eqnarray}
 \theta_1 = \log (\tau_1^2), \quad
 \theta_2 = \log (\tau_2^2), \quad
 \theta_3 = \log (\sigma^2) .
\end{eqnarray}

In this case, 
\begin{eqnarray}
 && \frac{\partial Q}{\partial\theta_1} 
    = \frac{\partial^2 Q}{\partial\theta_1\partial\theta_1} 
    = \left[ \begin{array}{cc} 
             \tau_1^2 & 0  \\
               0      & 0  \end{array}\right], \quad
  \frac{\partial Q}{\partial\theta_2} 
     = \frac{\partial^2 Q}{\partial\theta_2\partial\theta_2} 
     = \left[ \begin{array}{cc } 
               0 & 0   \\
               0 & \tau_2^2  \end{array}\right] \quad
  \frac{\partial Q}{\partial\theta_3} 
     = \frac{\partial^2 Q}{\partial\theta_2\partial\theta_2} 
     = \left[ \begin{array}{cc} 
               0 & 0  \\
               0 & 0  \end{array}\right] \nonumber \\
 && \frac{\partial^2 Q}{\partial\theta_i\partial\theta_j} 
     = \left[ \begin{array}{cc} 
               0 & 0   \\
               0 & 0 \end{array}\right]  \quad  (i\neq j), \quad
 \frac{\partial R}{\partial\theta_1} =  \frac{\partial R}{\partial\theta_2} = 0, \quad
 \frac{\partial R}{\partial\theta_3} =  \sigma^2, \\
&& \frac{\partial^2 R}{\partial\theta_3\partial\theta_3} = \sigma^2,  \quad 
   \frac{\partial^2 R}{\partial\theta_i\partial\theta_j} = 0\quad (\mbox{unless }i=j=3). \nonumber
\end{eqnarray}
 
Further, since $F$, $G$ and $H$ do not depend on $\theta$, 
$\displaystyle\frac{\partial F}{\partial\theta}= \frac{\partial^2 F}{\partial\theta\partial\theta^T}=0$, 
$\displaystyle\frac{\partial G}{\partial\theta}= \frac{\partial^2 G}{\partial\theta\partial\theta^T}=0$ and 
$\displaystyle\frac{\partial H}{\partial\theta}= \frac{\partial^2 H}{\partial\theta\partial\theta^T}=0$ hold,
where 0 indicates a zero matrix with appropriate size.

For Whard  data, the standard seasonal adjustment model with $m_1=2$, $m_2=1$ is estimated
using the initial estimates of parameters, 
$\theta = (\log\tau_1^2, \log\tau_2^2, \log\sigma^2) =(-9.21034, -10.81978, -8.5179)^T$.
The log-likelihood of the model with these initial parameters is $\ell (\theta )=-346.5115$
and the Gradient obtained by the differential filter are shown in the Table \ref{Tab_TS_model}.

The maximum likelihood estimate of the paramete vector is
$\hat\theta =(-12.10001,-10.04570,-9.85025)$ with maximum log-likelihood 
$\ell (\hat\theta )=384.9600$.
The gradient and the Hessian matrix of this model are shown in table.
The bias correction term of this estimated standard seasonal adjustment model is 3.9558.



\subsection{Seasonal adjustment model with stationary AR component}

The third example is a seasonal adjustment model with statinary AR component
\begin{eqnarray}
y_n = T_n + S_n + p_n +w_n,
\end{eqnarray}
where $T_n$ and $S_n$ are the trend component and the seasonal component
introduced in the previous subsection and $p_n$ is an AR component with AR order $m_3$ defined by
\begin{eqnarray}
 p_n = \sum_{j=1}^{m_3} a_j p_{n-j} + v_n^{(t)}. 
\end{eqnarray}
Here $v_n^{(t)}$is a Gaussian white noise with variance $\tau_3^2$.
The model contains $4+m_3$ parameters and the parameter vector is given by
$\theta =(\theta_1 ,\ldots ,\theta_{4+m_3})\equiv \tau_1^2,\tau_2^2,\tau_3^2,\sigma^2
 a_1,\cdots ,a_{m_3})^T$.

The matrices $F$, $G$, $H$, $Q$ and $R$ are defined by
{\setlength{\arraycolsep}{1mm}
\begin{eqnarray}
x_n = \left[ \begin{array}{c} 
             T_n     \\
             T_{n-1} \\ \hline
             S_n     \\
             S_{n-1} \\
             \vdots  \\ 
             S_{n-p+2} \\ \hline
             p_{n-1}  \\
             p_{n-2}  \\
             \vdots   \\
             p_{n-m_3} \end{array}\right],\quad
F &=& \left[ \begin{array}{cc|cccc|cccc} 
             2 &-1 &   &   &   &   \\
             1 & 1 &   &   &   &   \\
             \hline
               &   &-1 &-1 &\cdots&-1\\
               &   & 1 &   &   &   \\
               &   &   &\ddots&&   \\ 
               &   &   &   & 1 &   \\
             \hline
               &   &   &   &   &   &a_1 & a_2 &\cdots & a_{m_3} \\
               &   &   &   &   &   & 1  &     &       &    \\
               &   &   &   &   &   &    & \ddots &    &    \\
               &   &   &   &   &   &    &        & 1  &
         \end{array}\right],\quad
G = \left[ \begin{array}{ccc} 
             1 & 0 & 0  \\
             0 & 0 & 0 \\  \hline
             0 & 1 & 0 \\
             0 & 0 & 0 \\
          \vdots&\vdots&\vdots\\ 
             0 & 0 & 0 \\  \hline
             0 & 0 & 1 \\
             0 & 0 & 0 \\
          \vdots&\vdots&\vdots\\ 
             0 & 0 & 0          
 \end{array}\right] \\
H &=& [\begin{array}{cccccccccc} 1&0&1&0&\cdots &0&1&0&\cdots &0 \end{array}] \nonumber \\
Q &=& \left[ \begin{array}{ccc} 
             \tau_1^2 & 0        & 0  \\
               0      & \tau_2^2 & 0  \\
               0      & 0  & \tau_3^2\end{array}\right], \quad
R = \sigma^2.
\end{eqnarray}  
}

In this case, 
\begin{eqnarray}
 && \frac{\partial Q}{\partial\theta_1} = \frac{\partial^2 Q}{\partial\theta_1\partial\theta_1}
    = \left[ \begin{array}{ccc} 
             \tau_1^2 & 0 & 0  \\
               0      & 0 & 0  \\
               0      & 0 & 0  \end{array}\right], \quad
  \frac{\partial Q}{\partial\theta_2} = \frac{\partial^2 Q}{\partial\theta_2\partial\theta_2}
    =\left[ \begin{array}{ccc} 
               0 & 0  & 0 \\
               0 & \tau_2^2 & 0 \\
               0 & 0  & 0\end{array}\right], \quad
   \frac{\partial Q}{\partial\theta_3} = \frac{\partial^2 Q}{\partial\theta_3\partial\theta_3}
    =\left[ \begin{array}{ccc} 
               0 & 0  & 0  \\
               0 & 0  & 0  \\
               0 & 0  & \tau_3^2\end{array}\right], \nonumber\\
 && \frac{\partial^2 Q}{\partial\theta_i\partial\theta_j}
    =\left[ \begin{array}{ccc} 
               0 & 0  & 0  \\
               0 & 0  & 0  \\
               0 & 0  & 0  \end{array}\right], \quad \mbox{for }i \neq j \nonumber\\
 && \frac{\partial R}{\partial\theta_1} = 
    \frac{\partial R}{\partial\theta_2} = 0, \quad
    \frac{\partial R}{\partial\theta_3} = 
    \frac{\partial^2 R}{\partial\theta_3\partial\theta_3} = \sigma^2, \quad
    \frac{\partial^2 R}{\partial\theta_i\partial\theta_j} = 0 \quad (\mbox{unless }i,j=3)\\
&&\left(\frac{\partial F}{\partial \theta_k}\right)_{pq} =
   \left\{ \begin{array}{cl} 1  & \mbox{if } k=4+i, p=4, q=k, (i=1,\ldots ,m_3) . \\
             0                                      & \mbox{otherwise} \end{array}
   \right. \nonumber \\
&&\left(\frac{\partial^2 F}{\partial \theta_j \partial \theta_k}\right)_{pq} = 0
 \nonumber
\end{eqnarray}
where $\displaystyle\left(\frac{\partial F}{\partial \theta_k}\right)_{pq}$ and 
$\displaystyle\left(\frac{\partial^2 F}{\partial\theta_j \partial\theta_k}\right)_{pq}$
denote the $(p,q)$ components
of the matrices $\displaystyle\left(\frac{\partial F}{\partial \theta_k}\right)$ and 
$\displaystyle\left(\frac{\partial F^2}{\partial\theta_j \partial\theta_k}\right)$,
respectively.

For $m_1=2$, $m_2=1$ and $m_3=2$, the matrices $\displaystyle \frac{\partial F}{\partial \theta_j}, (j=5,6)$
are given by
\begin{eqnarray}
\frac{\partial F}{\partial \theta_5}
   = \left[ \begin{array}{ccc|cc}
            0 & \ldots & 0 & 0 & 0\\
            \vdots &\ddots & \vdots  & \vdots & \vdots\\
            0 & \ldots & 0 & 0 & 0\\
            \hline 
            0 & \ldots & 0 & 1 & 0\rule{0mm}{4mm} \\
            0 & \ldots & 0 & 0 & 0
            \end{array}  \right], \quad
\frac{\partial F}{\partial \theta_6}
   = \left[ \begin{array}{ccc|cc}
            0 & \ldots & 0 & 0 & 0\\
            \vdots &\ddots & \vdots  & \vdots & \vdots\\
            0 & \ldots & 0 & 0 & 0\\
            \hline 
            0 & \ldots & 0 & 0 & 1\rule{0mm}{4mm} \\
            0 & \ldots & 0 & 0 & 0
            \end{array}  \right].
\end{eqnarray}
%

\begin{table}[h]
\begin{center}
\caption{The gradient vectors and the Hessian matrix of the two TSAR models with $M_3=1$ obtained by
the proposed method.}\label{Tab_TSAR_model_1}
\vspace{2mm}
Maximum likelihood model\\
\begin{tabular}{crrrrr}
      & $\tau_1^2$ & $\tau_2^2$ & $\tau_3^3$ & $\sigma^2$ & $a_1^1 $  \\
\hline
          & $5.3908\!\times \!\!10^{-14}$ &  $5.4129\!\times \!\!10^{-5}$ &  $6.9080\!\times \!\!10^{-5}$ & $3.8147\!\times \!\!10^{-8}$ & 0.99990 \\
$\theta $ & -30.55150 & -9.82413 & -9.58025 & -17.0818 &  99.9900  \\
 \hline
$\frac{\partial\ell (\theta)}{\partial\theta}$& $-0.00000$ & $-0.00000$  & $-0.00112$ & $-0.00178$ &  0.00087   \\
\hline
     &  0.00000 & $-0.00000$ &  0.00000 & $-0.00000$ &   0.00000   \\
     &$-0.00000$&  30.99869  &  7.66932 &   0.01036 & $-0.07670$  \\
$\frac{\partial^2\ell (\theta)}{\partial\theta\partial\theta^T}$
     &  0.00000 &   7.66932  & 25.22303 &   0.00551 & $-0.24763$   \\
     &$-0.00000$ &  0.01036  &  0.00551 &  0.00178 &  $-0.00006$  \\
     &  0.00000 & $-0.07670$ & $-0.24763$ & $-0.00006$&   0.07058   \\
\hline
\end{tabular}
\vspace{2mm}

Local MLE model\\
\begin{tabular}{crrrrrr}
      & $\tau_1^2$ & $\tau_2^2$ & $\tau_3^3$ & $\sigma^2$ & $a_1^1 $  \\
\hline
          & $5.4741\!\times \!\!10^{-6}$ &  $4.3834\!\times \!\!10^{-5}$ &  $2.2564\!\times \!\!10^{-23}$ & $5.2261\!\times \!\!10^{-5}$ & 0.96816 \\
$\theta $ & -12.11548 & -10.03510 & -54.44829 & -9.85926 &  96.81629  \\
 \hline
$\frac{\partial\ell (\theta)}{\partial\theta}$& $0.00000$ & 0.00000 & $0.00000$ & $0.00000$ & 0.00000   \\
\hline
     &  8.79305 & $ 1.09359$ &  0.00000 &   3.43159 & $-0.00443$  \\
     &$ 1.09359$&  19.24777  &  0.00000 &  11.29246 & $ 0.00427$  \\
$\frac{\partial^2\ell (\theta)}{\partial\theta\partial\theta^T}$
     &  0.00000 &   0.00000  &$-0.00000$&   0.00000 &   0.00000   \\
     &  3.43159 & $11.29246$ &  0.00000 & $11.86720$&   0.00053  \\
     &$-0.00443$& $0.00427 $ &$0.00000$& $0.00053$&   0.00108  \\
\hline
\end{tabular}
\end{center}
\end{table}

Table \ref{Tab_TSAR_model_1} shows the gradients and the Hessians of the TSAR model with the 
first order AR component $m_3=1$.
Two models with the maximum likelihood estimates and the local maximum likelihood estimates (the 
second best model) are shown.

The maximum likelihood estimates of the parameter the AR coefficient is $a_1^1 = 0.9999$ 
that show the AR process has  of the model has almost unit root.
The log-likelihood of the maximum likelihood model is 392.1341.
Instead, the varainces of the trend component and observation noise are very small.
The logo-likelihood of the model is 385.2710.
On the other hand, the variance of the AR coefficient of the second best model is almost zero.
The bias correction terms are $b(GIC)=1.7655$ and 4.2173, respectively.

Table \ref{Tab_TSAR_model_2} shows the gradients and the Hessians of the TSAR model with the 
second order AR component $m_3=2$.
In this case the observation noise variance is almost zero.
The log-likelihood and the bias correction term are $\ell(\hat\theta )= 393.0525$ and $b(GIC)=4.1985$,
respectively.

\begin{table}[tbp]
\begin{center}
\caption{The gradient vectors and the Hessian matrix of TSAR model with $m_3=2$ obtained by
the proposed method.}\label{Tab_TSAR_model_2}
\vspace{2mm}

Optimized model:\\
\begin{tabular}{crrrrrr}
      & $\tau_1^2$ & $\tau_2^2$ & $\tau_3^3$ & $\sigma^2$ & $a_1^2 $ & $a_2^2$ \\
\hline
          & $2.1040\!\times \!\!10^{-19}$ &  $6.2681\!\times \!\!10^{-5}$ &  $3.4119\!\times \!\!10^{-5}$ & $8.7565\!\times \!\!10^{-27}$ &  $1.36666$ &  $-0.37573 $ \\
$\theta $ & -43.00526 & -9.67745 & -10.28566 & -34.06057 &  136.66603 & -37.57285 \\
\hline
$\frac{\partial\ell (\theta)}{\partial\theta}$& $-0.00000$ &  $ 0.00000$ & -0.00000 &  -0.00000  &   0.00000 & 0.00000 \\
 \hline
     &  0.00000 & $-0.00000$  &  0.00000 & $-0.00000$ &   0.00000  &  $ 0.0000$ \\
     &$-0.00000$&  39.50535  &  5.86945 &   0.00000  & $-0.09412$ &  $-0.08503$ \\
$\frac{\partial^2\ell (\theta)}{\partial\theta\partial\theta^T}$
     &  0.00000 &   5.869465  & 20.36345 &   0.00000 &   0.01534  &  $-0.39764$ \\
     &$-0.00000$ &   0.00000  &  0.00000 & $ 0.00000$&  $-0.00000$  &  $-0.00000$ \\
     &  0.00000  & $-0.09412$ &  0.01534 & $-0.00000$&  0.22332  &  $0.21832$  \\
     &$ 0.00000$ & $-0.08503$ &$-0.39764$&$-0.00000$  &  0.21832 &   0.22488  \\
\hline
\end{tabular}
\end{center}
\end{table}

Table \ref{Tab_TSAR_model_3} shows the log-likelihood, the number of parameters, 
bias correction term of BIC, AIC and GIC of various state-space models, sucha as
the trend model with order 1 and 2 ($m_1=1$ or 2, $m_2=1, m_3=0$, the 
seasonal adjustment model with AR order $m_3=0,1,2,3$.
It can be seen that the $b_{AIC}$ and $b_{GIC}$ are considerablly different,
in this case the both criteria select the seme model $m_1=2$, $m_2=1$ and
$m_3=1$.

\begin{table}[tbp]
\begin{center}
\caption{Log-likelihoods and bias correction terms of AIC and GIC for the seasonal
adjustment model with AR components.}\label{Tab_TSAR_model_3}
\vspace{2mm}
\begin{tabular}{cccccccc}
 \hline
   $m_1$ & $m_2$ & $m_3$ & log-likelihood  &  $b_{\mbox{\small AIC}}$&  $b_{\mbox{\small GIC}}$ & AIC & GIC  \\
\hline
 1 & 0 & 0 &   319.5067  &   2 &    1.4669  & $-635.0134$ & $-636.0796$ \\
 2 & 0 & 1 &   296.7171  &   2 &    1.9232  & $-589.4342$ & $-589.5878$ \\
 2 & 1 & 0 &   384.9600  &   3 &    3.9558  & $-763.9201$ & $-762.0084$ \\
 2 & 1 & 1 &   392.1015  &   5 &    1.8232  & $-774.2030$ & $-780.5566$ \\
 2 & 1 & 2 &   393.0525  &   6 &    4.1998  & $-774.1050$ & $-777.7054$ \\
 2 & 1 & 3 &   393.1091  &   7 &    4.4603  & $-772.2182$ & $-777.2976$ \\
\hline
\end{tabular}
\end{center}
\end{table}


\section{Summary}
The gradient and Hessian of the log-likelihood of linear state-space 
model are given. 
Details of the implementation of the algorithm for standard seasonal adjustment model,
seasonal adjustment model with stationary AR component and ARMA model are given.
For each implementation, comparison with a numerical difference method is shown.

\vspace{15mm}
\noindent{\Large\bf Aknowledgements}

This work was supported in part by JSPS KAKENHI Grant Number 18H03210.
The author is grateful to the project members, Prof. Kunitomo, Prof Nakano,
Prof. Kyo, Prof. Sato, Prof. Tanokura and Prof. Nagao for their 
stimulating discussions.

\end{document}